\documentclass[twocolumn,amsmath,amssymb, prd, superscript address,preprintnumbers]{revtex4}

\usepackage{graphicx}
\usepackage{dcolumn}
\usepackage{bm}
\usepackage{epstopdf}%

\begin{document}

\preprint{MIT-CTP-3687}

\title{Example of a Hidden Flavor Sector}
\author{Brian Patt}
\affiliation{Center for Theoretical Physics, Laboratory for Nuclear Science and Department of Physics, Massachusetts Institute of Technology, Cambridge, Massachusetts 02139, USA}
\author{David Tucker-Smith}
\affiliation{Department of Physics, Williams College, Williamstown, Massachusetts 01267, USA}
\author{Frank Wilczek}
 \affiliation{Center for Theoretical Physics, Laboratory for Nuclear Science and Department of Physics, Massachusetts Institute of Technology, Cambridge, Massachusetts 02139, USA}

\date{\today}

\begin{abstract}

We exemplify earlier general considerations on flavor symmetry breaking employing a hidden sector and exploiting supersymmetry in a specific model.   The model is at best a caricature of reality, but it is sufficient to display mechanisms for the natural emergence of minimal low-energy Higgs structure, non-trivial texture zeros and symmetry, and coupling to the standard model only through Higgs fields, which could be considerably more general.  

\end{abstract}

\maketitle


Quark and lepton masses and mixings are accommodated, rather than explained, within the standard model.   There have been many attempts to explain perceived patterns in their observed values, including some fairly economical, attractive proposals based on different sorts of approximate, or spontaneously broken, family symmetry.  Unfortunately the relationship between the hypothesized symmetry and practical observables is generally not very direct, because the practical observables are complicated combinations of the natural objects for theoretical consideration.    The natural objects for theoretical consideration are coupling coefficients in the Lagrangian that governs the unbroken symmetry phase, while the observed masses and mixings reflect the eigenvalues and orientation of the eigenvectors of the mass matrix which emerges after symmetry breaking.   In a sense this complication is welcome, because -- with one exception -- no simple pattern is manifest in the practical observables.   The exception is the $\theta$ parameter of QCD, which is essentially the phase of the determinant of the quark mass matrix.  $\theta$ is observed to be very small ($|\theta | \lesssim 10^{-9}$), and this smallness can be explained robustly through the hypothesis of an appropriate approximate (asymptotic and spontaneously broken) $U(1)$ symmetry, as first proposed by Peccei and Quinn \cite{PecceiQuinn}.   

These considerations, that is the attractiveness of the general idea of family symmetry and the difficulty of distinguishing among specific models, motivated us in \cite{firstPaper} to consider two broad hypotheses concerning family symmetry and its breaking, that permit some interesting general conclusions, independent of their detailed implementation.  The first broad hypothesis is that the Higgs doublet(s) of the (supersymmetric) standard model transforms non-trivially under a large family symmetry that is broken at a very large mass scale $f$ (e.g., $f \gtrsim 10^{16}$ GeV).  (These ideas or closely related ones have also been considered, {\it e.g.} in \cite{LightHiggsFam,HeavyHiggsFam}.)  The family symmetry must therefore commute with the standard model $SU(3)\times SU(2)\times U(1)$ gauge symmetry.  One can have  -- and we need -- a hidden sector of fields that transform under the family symmetry but are $SU(3)\times SU(2)\times U(1)$ singlets.  The ``standard model'' (i.e., light, with mass $\ll f$) Higgs fields appear within a multiplet of $SU(2)\times U(1)$ doublets, most of which acquire mass of order $f$.   The couplings of the standard model Higgs fields, and thereby ultimately quark and lepton masses and mixings, are determined by the equations which pick out the light components of the multiplet.  

The emergence of light, non-derivatively coupled fields suitable to play the role of standard model Higgs fields is natural only within the context of supersymmetry.   Some form of approximate, low-energy supersymmetry is also desirable for insuring quantitative unification of couplings, and for other reasons \cite{UnificationSUSY}.   This motivates us to add low-energy supersymmetry as our second broad hypothesis.   In \cite{firstPaper} we showed how the extra structure associated with family symmetry breaking in a hidden sector and supersymmetry can naturally engender a pair of Higgs doublets, as required in the minimal supersymmetric standard model.   We also discussed possible advantages of this framework for addressing the $\mu$, B$\mu$, and approximate flavor universality problems \cite{SUSYFlavor,Polonsky:2001pn} in supersymmetric model-building, and described a natural mechanism for generating $\mu$ that requires couplings between hidden-flavor-sector states  and the light Higgs doublets.

In this note we will exemplify two other possible consequences of our hypotheses that were mentioned in \cite{firstPaper}, but did not appear in the models presented there.  The first is the possibility of deriving desirable textures (e.g., zeros and symmetries in the mass matrix).   The second is the possible emergence of light moduli fields from the hidden flavor sector that have unsuppressed coupling to standard model Higgs fields.

{\it Multiplets and Symmetry Breaking}:  The general structure of our models is as follows.  All our fields are chiral superfields.   Family breaking fields, denoted by letters around $T$, are standard model singlets, some of which will acquire vacuum expectation values (VEVs) at a large scale.  Generalized Higgs fields with $SU(3)\times SU(2) \times U(1)$ quantum numbers $(1, 2, +\frac{1}{2})$ will be denoted $U$, and generalized Higgs fields with $SU(3)\times SU(2) \times U(1)$ quantum numbers $(1, 2, -\frac{1}{2})$ will be denoted $D$.  The family sector gives
masses to the generalized Higgs fields through couplings of the form $U D T$. These couplings will leave
one component from each of $U$ and $D$ massless, and these components play the role of the 
standard model Higgs fields.  Terms of the form $q U u^c$ will give masses
to the up quarks, where $q$ and $u$ are quark fields, where of course $q$ represents the $SU(2)$ doublets of left-handed quark fields and $u^c$ the right-handed hypercharge $-\frac{2}{3}$ quark fields.  More precisely, it is couplings to the (approximately) massless component of $U$ that give rise to mass, and the form of the up-quark mass matrix will be determined by the form of that component.  Similar
terms involving $D$ give mass to down quarks and leptons.

In this paper we will consider a model with two generations and $SU(2)$ family symmetry.  All the standard model fermion
fields transform normally, but carry one additional family $SU(2)$ index.  
For example the field ${u^c}_{k}$ has the $SU(2)$ family index $k$ and
therefore represents both the up quark and charm quark.
The generalized Higgs fields form a pair of family spin one fields with 
appropriate gauge charges for the Higgs.  We denote the generalized 
Higgs fields as ${U}_{1m}^{jk}, {U}_{2m}^{jk}, {D}_{1m}^{jk}, {D}_{2m}^{jk}$,
where $m$ is the $SU(2)_{weak}$ index. 
Under $SU(2)_{fam}$ each family spin one field is a symmetric rank 2 
tensor, with indices $ij$.  The family breaking fields are a set of 
five flavor spin one fields, denoted $X_{ij}, T_{ij}, S_{ij}, R_{ij}, P_{ij}$.
Finally we also introduce three singlet fields, $\eta$, $\xi$, and $\zeta$.

We introduce four $U(1)$ global symmetries, each associated with one of the generalized Higgs fields.  The following table exhibits our $U(1)$ charge 
assignments:
\begin{displaymath}
\begin{array}{*{5}c}
\rm{field}         & U(1)_E & U(1)_F & U(1)_G & U(1)_H \\
U_1    &      1 &      0 &      0 &      0 \\
U_2    &      0 &      1 &      0 &      0 \\
D_1    &      0 &      0 &      1 &      0 \\
D_2    &      0 &      0 &      0 &      1 \\
Q_{i,j}        &      0 &      0 &      0 &      0 \\
{u}^c_k      &      0 &     -1 &      0 &      0 \\
{d}^c_l      &      0 &      0 &      0 &     -1 \\
X_{ij}         &      0 &      0 &      0 &      0 \\
T_{ij}         &     -1 &      0 &     -1 &      0 \\
S_{ij}         &      0 &     -1 &     -1 &      0 \\
R_{ij}         &     -1 &      0 &      0 &     -1 \\
P_{ij}         &      0 &     -1 &      0 &     -1 \\
\eta           &      0 &      0 &      0 &      0 \\
\xi            &      0 &      1 &      1 &      0 \\
\zeta          &      1 &      0 &      0 &      1 \\
\end{array} 
\end{displaymath}

We impose a $U(1)_R$
symmetry under which every field has an R-charge of $2/3$.
With this, we have all the necessary information to construct the superpotential.
The permissible terms are then as follows:

\newcounter{termset3}
\begin{list}{\arabic{termset3})}{\usecounter{termset3}}
\item Terms that couple the Higgs to the quarks. 
\begin{equation*}
W_{\rm quarks} = \lambda_U \epsilon^{im} q_{i,j} {U_2}_{m}^{jk} {u}^c_k +
 \lambda_D \epsilon^{im} q_{i,j} D_{2m}^{j,l} {d}^c_l \;\;\;\;\;\;\; 
\end{equation*}
\item Terms that couple the Higgs to the family breaking fields.
\begin{eqnarray*}
\;\;\;\;
W_{\rm Higgs} & = & \lambda_T \epsilon^{mn} \epsilon_{kl}
{U}_{1m}^{ik} {D}_{1n}^{jl} T_{ij} 
+ \lambda_S \epsilon^{mn} \epsilon_{kl} {U}_{2m}^{ik} {D}_{1n}^{jl} S_{ij} \\
& + & \lambda_R \epsilon^{mn} \epsilon_{kl} {U}_{1m}^{ik} {D}_{2n}^{jl} R_{ij} 
+ \lambda_P \epsilon^{mn} \epsilon_{kl} {U}_{2m}^{ik} {D}_{2n}^{jl} P_{ij} \\
\end{eqnarray*}
\item Terms that make a potential for the family breaking fields and scalars.
\begin{eqnarray*}
W_{\rm potential} & = & \lambda_{\eta} \eta \epsilon^{ac}\epsilon^{bd}X_{ab}X_{cd} +
\lambda_{\xi} \xi \epsilon^{ac}\epsilon^{bd}X_{ab}S_{cd} \\
& + & \lambda_{\zeta} \zeta \epsilon^{ac}\epsilon^{bd}X_{ab}R_{cd} 
\end{eqnarray*}
\end{list}

Now let us determine the moduli space of this model's family breaking 
sector.  
We have to compute the singlet F-terms, in the case for $\eta$, $\xi$, and 
$\zeta$.  It proves convenient for $SU(2)$ spin one family fields to use $SO(3)$ vector
notation, so we can find the moduli space and
the form of the light Higgs mode.  These vectors will be complex, and we use 
arrows to denote them, i.e. $\vec{X}$ 
corresponds
to $X_{ij}$.
In $SO(3)$ notation the superpotential reads:
\begin{equation*}
W_{\rm potential} =
2\lambda_{\eta} \eta \vec{X} \cdot \vec{X} +
2\lambda_{\xi} \xi \vec{X} \cdot \vec{S} +
2\lambda_{\zeta} \zeta \vec{X} \cdot \vec{R}
\end{equation*}
Thus the $\eta$ F-term vanishes for:
\begin{equation*}
2\lambda_{\eta} \eta \vec{X} \cdot \vec{X} = 0
\end{equation*}
The solution of this equation is any vector whose real part is orthogonal to 
its imaginary part, and whose
real part has the same magnitude as its imaginary part.  This
vector can be simplified by rotating to a frame where the real part is
in the positive $x$ direction, and then by rotating about the $x$-axis 
such that the imaginary 
part is in the negative $y$ direction. After these rotations
the solutions for $\vec{X}$ take the form
$\vec{X} = (\alpha, -i\alpha, 0)^t$.  
Now the $\xi$ F-term gives the constraint $2\lambda_{\xi} \xi \vec{X} \cdot \vec{S} = 0$, which
implies that the form of $\vec{S}$ is
$\vec{S} = (s_1, -i s_1, s_3)^t$.  
Similarly, the $\zeta$ F-term implies $\vec{R} = (r_1 , -i r_1 , r_3)^t$.  

At this level there are no constraints on $T_{ij}$; every component
of $T_{ij}$ is a flat direction.  Therefore $T_{ij}$ has a completely arbitrary
VEV, $\vec{T} = (t_1, t_2, t_3)^t$.  
Similar remarks apply to $P$, but we assume that the vacuum expectation value of
$P_{ij}$ is zero, as could be insured by 
small supersymmetry breaking terms in the potential.  For our model it will be
important to assume that the box product, $(\vec{T} \times \vec{R}) \cdot \vec{S}$, does not vanish, 
which is generically true.

Now let us determine which modes of the generalized Higgs fields are massless.  For this, it is convenient to convert the Higgs superpotential into $SO(3)$ notation.
Using a prime to distinguish tensors with upper indices, such as $X^{ij}$,  we have 
\begin{eqnarray*}
W_{\rm Higgs} =  
2 \lambda_X ({\vec{T}}^{\prime})^{*}\cdot 
(\vec{U}_1 \times \vec{D}_1) +
2 \lambda_S ({\vec{S}}^{\prime})^{*}\cdot 
(\vec{U}_2 \times \vec{D}_1) &&\\
\;\;\;\;\;\;\;\;\;\;\;\;+\;
2 \lambda_R ({\vec{R}}^{\prime})^{*}\cdot 
(\vec{U}_1 \times \vec{D}_2) +
2 \lambda_P ({\vec{P}}^{\prime})^{*}\cdot 
(\vec{U}_2 \times \vec{D}_2) && \\
\end{eqnarray*}
To identify the light $U$-type Higgs field, first use vector identities to
rewrite the superpotential as:
\begin{eqnarray*}
W_{\rm Higgs} =   
2 \lambda_T \vec{D}_1 \cdot \left[ ({\vec{T}}^{\prime})^{*} 
\times \vec{U}_1\right] +
2 \lambda_S \vec{D}_1 \cdot \left[ ({\vec{S}}^{\prime})^{*} 
\times \vec{U}_2\right] && \\ 
\;\;\;\;\;\;\;\;\;\;\;\;+\;
2 \lambda_R \vec{D}_2 \cdot \left[ ({\vec{R}}^{\prime})^{*}
\times \vec{U}_1\right] +
2 \lambda_P \vec{D}_2 \cdot \left[ ({\vec{P}}^{\prime})^{*}
\times \vec{U}_2\right] &&\\ 
\end{eqnarray*}
Using this, we readily derive the down Higgs F-terms in the compact form:
\begin{eqnarray}
2 \lambda_T ({\vec{T}}^{\prime})^{*} \times \vec{U}_1 +  
2 \lambda_S ({\vec{S}}^{\prime})^{*} \times \vec{U}_2 & = & \vec{0} \\
2 \lambda_R ({\vec{R}}^{\prime})^{*} \times \vec{U}_1 +
2 \lambda_P ({\vec{P}}^{\prime})^{*} \times \vec{U}_2 & = & \vec{0}
\end{eqnarray}
The massless up Higgs modes are any modes which satisfy the above
constraints, taking into account the VEVs of $\vec{T}^{\prime}$,
$\vec{S}^{\prime}$, $\vec{R}^{\prime}$ (but $\vec{P}^{\prime} =0$).   The 
the second constraint implies that  $\vec{U}_1 = C_1 ({\vec{R}}^{\prime})^{*}$
where $C_1$ is a constant.
Now we substitute this into the first constraint and find
\begin{equation*}
2 C_1 \lambda_T ({\vec{T}}^{\prime})^{*} \times
 ({\vec{R}}^{\prime})^{*}   +  
2 \lambda_S ({\vec{S}}^{\prime})^{*} \times \vec{U}_2 = \vec{0}
\end{equation*}
This implies that the cross product of the two terms vanishes, so that
\begin{equation*}
\left[({\vec{T}}^{\prime})^{*} \times ({\vec{R}}^{\prime})^{*}\right] \times
\left[({\vec{S}}^{\prime})^{*} \times \vec{U}_2\right] = \vec{0}
\end{equation*}
Finally, using the identity
$\vec{A}\times(\vec{B}\times\vec{C})=(\vec{A}\cdot\vec{C})\vec{B}-
(\vec{A}\cdot\vec{B})\vec{C}$, 
\begin{equation*}
\left[ \vec{A}
\cdot \vec{U}_2 \right] ({\vec{S}}^{\prime})^{*} 
-\left[ \vec{A}\cdot ({\vec{S}}^{\prime})^{*}\right]
\vec{U}_2
=
\vec{0}
\end{equation*}
where $\vec{A} \equiv ({\vec{T}}^{\prime})^{*} \times ({\vec{R}}^{\prime})^{*}$.
Since the coefficient of $\vec{U}_2$ does not vanish, this equation implies that $({\vec{S}}^{\prime})^{*}$ is parallel to
$\vec{U}_2$.
Thus the second term of
equation 1 vanishes, and so must the first term.  Since $({\vec{T}}^{\prime})^{*}$ and $({\vec{R}}^{\prime})^{*}$ are not
parallel, this can only happen if $C_1=0$.

Putting it all together, we find that the light $U$-type Higgs field is of the form
$\vec{U}_1 = 0, \vec{U}_2 = ({s_1}, i {s_1}, {s_3})^t$,
or in the $SU(2)$ spin one tensor notation:
\begin{eqnarray*}
U_1 = 
\left( \begin{array}{*{2}c}
0 & 0 \\
0 & 0 \\
\end{array} \right) & \;\;\;\;\;\;\;\;\; &
U_2 = 
\left( \begin{array}{*{2}c}
0     &  -s_3    \\
- s_3 & -2 {s_1} \\
\end{array} \right) \equiv \left( \begin{array}{*{2}c}
0     &  -\gamma    \\
-\gamma & \beta \\
\end{array} \right)
\end{eqnarray*}
We can repeat all this analysis to find the form of the massless $D$-type Higgs field
\begin{eqnarray*}
D_1  =  
\left( \begin{array}{*{2}c}
0 & 0 \\
0 & 0 \\
\end{array} \right) & \;\;\;\;\;\;\;\;\; &
D_2 =  
\left( \begin{array}{*{2}c}
0     & - r_3   \\
-r_3  & -2{r_1} \\
\end{array} \right) \equiv \left( \begin{array}{*{2}c}
0     & -\tau \\
-\tau & \omega \\
\end{array} \right) 
\end{eqnarray*}

{\it Mass Matrix, Texture Relation}: It is now straightforward to calculate the Yukawa couplings between the
quark fields and the light Higgs fields, and thus derive the mass matrices.  
In view of the normalized forms
\begin{equation*}
U_2= 
\frac{1}{\sqrt{2\gamma^2+\beta^2}}
\left( \begin{array}{*{2}c}
0       & -\gamma \\
-\gamma &   \beta \\
\end{array} \right) \tilde{U}_{\rm light}
\end{equation*}
\begin{equation*}
D_2 = 
\frac{1}{\sqrt{2\tau^2+\omega^2}}
\left( \begin{array}{*{2}c}
0     & -\tau \\
-\tau &  \omega \\
\end{array} \right) \tilde{D}_{\rm light}
\end{equation*}
we simply plug them into the quark-Higgs coupling terms of the superpotential, to derive up- and down-type quark mass matrices of the forms
\begin{eqnarray*}
\frac{\lambda_U}{\sqrt{2{\gamma}^2+{\beta}^2}}
\left( \begin{array}{*{2}c}
      0       &  -\gamma \\
      -\gamma &  \beta  \\
\end{array} \right)  \\
\frac{\lambda_D}{\sqrt{2{\tau}^2+{\omega}^2}}
\left( \begin{array}{*{2}c}
0      & -\tau  \\
-\tau  & \omega \\
\end{array} \right) \\
\end{eqnarray*}

As is well known \cite{CabMassRel}, this texture leads to a successful relation among the quark masses and the Cabibbo angle, of the form
\begin{equation*}
\theta_{\rm{Cabibbo}} \approx \frac{\tau}{\omega}-\frac{\gamma}{\beta} \approx \sqrt{\frac{m_d}{m_s}} - \sqrt{\frac{m_u}{m_c}}
\end{equation*}

In the present framework, the quark mass hierarchy actually results from a 
hierarchy between the vacuum expectation values in the flat directions.   To the extent that these are arbitrary, we have the possibility within inflationary cosmology that they might assume different values in different parts of the Multiverse.  This opens up the possibility that selection effects could operate to determine the observed values of the quark masses and mixing angles.   

{\it Moduli Coupling}:  As we mentioned near the start, low-mass dimension interaction terms of the type 
$\lambda_{mod} U_{\rm light}^{\dagger}U_{\rm light} F^{\star}F$, where 
$F$ is a moduli field are not forbidden by general principles (here we are using ordinary, not superfield, notation).  We
want to explore when these terms will actually arise from our models.

Schematically, the superpotential terms from which such interactions might arise are of the form $UDT$.
These coupling terms contain many heavy modes and some light modes, but we are interested in the interactions involving light modes only. In order to isolate the 
interactions of the light fields, we must integrate out the heavy modes.

In our model the non-trivial vacuum expectation values of the moduli fields provide masses for the generalized Higgs fields.
This implies that the low energy effective Lagrangian
has no coupling between quanta of the moduli fields and the light Higgs fields, because variations of the moduli fields do not contribute to the light Higgs field mass.   However, moduli such as $P_{ij}$, which we 
assumed to have zero vacuum expectation value, can interact with the light Higgs field.
One method to see if they do interact, is to notionally assign
these moduli fields vacuum expectation values, and observe whether the light Higgs field remains massless. If it 
does not, then there must be a coupling after integrating out 
the heavy fields, because it is precisely through that coupling that the light 
Higgs field gets a mass.  In the model analyzed here, if $P_{ij}$ were to have a nonzero vacuum expectation value, then the
light Higgs field would become massive. Hence the light Higgs field 
couples to $P_{ij}$.  All these assertions can be checked by direct calculation.

{\it Conclusions}: We have demonstrated by example how the hypotheses that generalized Higgs fields transform non-trivially under family symmetry broken at a very high scale, and of low-energy supersymmetry, support natural mechanisms for the emergence of a minimal set of standard model Higgs fields, coupled to quarks with interesting textures.   In this framework, it is possible to have light moduli fields with unsuppressed couplings to the standard model Higgs sector.   Interesting, but because it relies on the accurate flatness of moduli potentials highly conjectural in application, is the feature that masses and mixings are effectively free parameters, and therefore open to selection effects.   Clearly it will be of considerable interest to extend our considerations to models with three families, and to unified gauge theories.   

{\it Acknowledgments:}  The work of DTS was supported in part by Research Corporation. This work is supported in part by funds provided by the U.S. Department of Energy (D.O.E.) under cooperative research agreement DE FG02-05 ER41360.

\end{document}